\documentclass[floats,floatfix,showpacs,amssymb,prd,twocolumn,superscriptaddress,nofootinbib,nolongbibliography,reprint]{revtex4-2}

\usepackage{amssymb,amsmath,verbatim,mathtools,needspace,enumitem,etoolbox,graphicx,physics,microtype,afterpage,xspace,tabularx,lmodern,multirow}
\usepackage{gensymb}
\usepackage[normalem]{ulem}
\usepackage[dvipsnames, usenames]{xcolor}
\definecolor{linkcolor}{rgb}{0.0,0.3,0.5}
\usepackage[unicode, colorlinks=true, linkcolor=linkcolor, citecolor=linkcolor, filecolor=linkcolor, urlcolor=linkcolor, linktocpage, breaklinks]{hyperref}
\usepackage[all]{hypcap}
\usepackage[T1]{fontenc}
\usepackage[utf8]{inputenc}
\usepackage[usenames,dvipsnames]{xcolor}
\usepackage{aas_macros}
\usepackage{makecell}
\usepackage{soul}
\usepackage{booktabs}
\usepackage{orcidlink}

\newcommand{\jhu}{\affiliation{William H. Miller III Department of Physics and Astronomy, Johns Hopkins University, 3400 North Charles
Street, Baltimore, Maryland 21218, USA}}
\newcommand{\milan}{\affiliation{Dipartimento di Fisica ``G. Occhialini'', Universit\'a degli Studi di Milano-Bicocca, Piazza della Scienza 3, 20126 Milano, Italy}}
\newcommand{\infn}{\affiliation{INFN, Sezione di Milano-Bicocca, Piazza della Scienza 3, 20126 Milano, Italy}}

\begin{document}
	
\title{Minimum gas mass accreted by spinning \\ intermediate-mass black holes in stellar clusters}

\begin{abstract}
 The spin of intermediate-mass black holes (IMBHs) growing through repeated black hole mergers in stellar clusters statistically asymptotes to zero. Putative observations of IMBHs with dimensionless spin parameter $\chi\gtrsim 0.6$ would require a phase of coherent gas accretion to spin up the black hole. We estimate the amount of gas necessary to produce a given IMBH spin. If the observed IMBH mass and spin are $M\gtrsim 1000M_\odot$ and $\chi\gtrsim 0.6$, respectively, the IMBH must have coherently accreted at least $\sim 100M_\odot$ of gas. In this scenario, as long as the spin is not maximal, the IMBH can only accrete half of its mass at most. Our estimates can constrain the relative contribution of accretion and mergers to the growth of IMBHs in dense stellar environments.
\vspace{1cm}
\end{abstract}

\author{Konstantinos Kritos$\,$\orcidlink{0000-0002-0212-3472}}
\email{kkritos1@jhu.edu}
\jhu

\author{Luca Reali$\,$\orcidlink{0000-0002-8143-6767}}
\email{lreali1@jhu.edu}
\jhu

\author{Davide Gerosa$\,$\orcidlink{0000-0002-0933-3579}}
\email{davide.gerosa@unimib.it}
\milan \infn

\author{Emanuele Berti$\,$\orcidlink{0000-0003-0751-5130}}
\email{berti@jhu.edu}
\jhu

\date{\today}
\maketitle

\section{Introduction}
\label{sec:Introduction}

Intermediate-mass black holes (IMBHs) with masses in the range $10^{2}$--$10^{6}M_\odot$ are hypothesized to bridge the gap between stellar-mass and supermassive 
black holes (BHs)~\cite{Miller:2003sc,2017IJMPD..2630021M,2020ARA&A..58..257G}. 
Despite the large number of IMBH candidates, conclusively confirming their nature and measuring their spin
remains challenging, and the mass estimates are also quite uncertain. 

In the supermassive BH regime, spin measurements via x-ray reflection spectroscopy point toward a population of objects with dimensionless spin magnitude $\chi=cJ/(GM^2)$ (where $M$ and $J$ are the BH mass and angular momentum, respectively) close to the extremal limit $\chi\simeq 1$.
However, systematic effects and selection biases for this technique still pose an issue~\cite{Reynolds:2020jwt}. 
Gravitational wave (GW) observations provide an alternative method for measuring the spins of merging BHs and extracting their properties. 
Current observations of merging stellar-mass BHs by LIGO-Virgo-KAGRA hint at a population of slowly spinning BHs~\cite{KAGRA:2021duu} (but see, e.g., Refs.~\cite{Roulet:2021hcu,Callister:2022qwb} for caveats).

Dense star clusters are ideal systems for producing IMBHs, and a few candidates have already been observed~\cite{2012ApJ...747L..13F,Huang:2024gpv,2024Natur.631..285H}.
The proposed formation channels for IMBHs in these environments include runaway stellar collisions eventually leading to the direct collapse of a single star~\cite{PortegiesZwart:2002iks,Prieto:2024pkt,Purohit:2024zkl,Fujii:2024uon}, repeated tidal disruption events~\cite{Stone:2016ryd,Rizzuto:2022fdp}, and hierarchical BH mergers~\cite{Miller:2001ez,Antonini:2018auk,Fragione:2021nhb,2021NatAs...5..749G,Atallah:2022toy}. 
While the spin of a BH formed from direct collapse of a star with mass $\gtrsim260M_\odot$ is highly uncertain~\cite{Heger:2002by}, the evolution of the spin from the other two channels (i.e., growth via accretion of stars or BHs) can be more easily predicted~\cite{Hughes:2002ei,Berti:2008af,Gerosa:2021hsc,Zevin:2022bfa}.

The dimensionless spin $\chi$ of a BH growing by episodic accretion or hierarchical mergers decreases with time and asymptotes to zero as long as the accreted mass is much smaller than the primary mass (i.e., in the so-called test-mass limit)~\cite{King:2008au,Hughes:2002ei}.
This well-known result stems from the fact that the innermost stable circular orbit (ISCO) radius for retrograde orbits in the symmetry plane of a spinning BH is larger than for prograde orbits.
If the accretion direction is random, the angular momentum subtracted by mass elements accreted on retrograde orbits is statistically larger than the angular momentum added by mass elements on prograde orbits, and therefore $\chi$ decreases with time.
On the other hand, if the BH merges with another BH of similar mass, the remnant spin magnitude jumps to $\chi\sim0.7$, with some scatter around that value depending on the progenitor binary mass ratio and spins~\cite{Berti:2008af,2017PhRvD..95l4046G,Fishbach:2017dwv,Baibhav:2020xdf}.

Reference~\cite{Gerosa:2021hsc} identifies an exclusion region beyond $M\gtrsim 50M_\odot$ and $\chi \lesssim0.2$, which successive BH mergers cannot efficiently contaminate. Alternative mechanisms are needed to explain BHs found in this exclusion region.
This work mainly focuses on another region avoided by hierarchical mergers, namely the high-mass, high-spin regime described in Sec.~\ref{sec:Black_hole_mass-spin_evolution}.
We propose that a single phase of coherent gas accretion may spin up BHs growing via hierarchical mergers, and in Sec.~\ref{sec:Distribution_of_accreted_gas_mass}, we estimate the distribution of the accreted gas mass.
Our method assumes knowledge of the IMBH's mass and spin, and it is agnostic to the technique used to measure these parameters. 
In Sec.~\ref{sec:Conclusions} we present our conclusions and the astrophysical implications of our results.

\section{Black hole mass and spin evolution}
\label{sec:Black_hole_mass-spin_evolution}

In this section, we describe our methodology to evolve the mass and spin of growing BHs through repeated mergers (Sec.~\ref{sec:Repeated_black_hole_mergers}) and coherent gas accretion (Sec.~\ref{sec:Coherent_gas_accretion}). In Sec.~\ref{sec:Minimum_required_accreted} we apply these methods to compute the minimum amount of gas that must be accreted to explain hypothetical observations of highly spinning IMBHs.

\subsection{Repeated black hole mergers}
\label{sec:Repeated_black_hole_mergers}

In the cores of dense stellar clusters, IMBHs can be assembled through successive BH mergers.
The heaviest stars of the cluster collapse into first-generation BHs, rapidly forming a BH subsystem in the cluster's core~\cite{Banerjee:2009hs}.
For a given stellar evolution model, the mass spectrum of first-generation BHs can be estimated depending on the initial mass function.
Detailed simulations are necessary to generate the time evolution of the cluster and the assembly of the IMBH using a star cluster evolution code~\cite{2013MNRAS.431.2184G,Antonini:2019ulv,Sedda:2021vjh,2022ApJS..258...22R,Kritos:2022ggc}.
The time information encoded in these simulations is unnecessary for our present purposes because we only care about the mass-spin evolution. 
The only nontrivial effect would be the time evolution of the BH mass function:
for example, due to mass segregation effects in the cluster, the heaviest BHs interact first and eject each other, leading to a depletion of the heaviest tail in the BH mass spectrum. 

We model the BH mass spectrum as a power law distribution  $p(m|m_{\rm min},m_{\rm max},\alpha)\propto m^{\alpha}$ in the range $m\in [m_{\rm min}, m_{\rm max}]$.
This simple assumption is motivated by the fact that the relation between the initial stellar mass and the final BH mass is uncertain. Detailed features in the spectrum itself have little effect on the growth of the IMBH, except for broad characteristics which are captured by the three parameters $m_{\rm min}$, $m_{\rm max}$, and $\alpha$ of our simplified model~\cite{Kritos:2022ggc}.

We also assume that the environment required to assemble the observed IMBH mass through repeated BH mergers and a single subsequent phase of coherent gas accretion can be realized in nature. In particular, when we simulate the growth of the BH through mergers, we do not compare the GW recoil velocity of the merger remnant with the escape velocity of the cluster: if the merger product were ejected, then it could not have produced the properties of the observed IMBH \cite{Gerosa:2019zmo}. Simulations suggest that IMBHs of up to $10^5M_\odot$ can be formed through this channel in systems with an escape velocity exceeding $\approx300\,\rm km\, s^{-1}$~\cite{Antonini:2018auk,Kritos:2022non}, and thus, in our analysis, we do not consider IMBHs heavier than this value.

We choose $\alpha=-2$ as a fiducial value, corresponding to equal BH mass per logarithmic mass bin~\cite{2019ARA&A..57..227K}, but we will also explore the effect of assuming $\alpha=-1$.
We consider a minimum BH mass $m_{\rm min}=5M_\odot$ following Ref.~\cite{KAGRA:2021duu} (but see e.g.~Ref.~\cite{Belczynski:2011bn} for uncertainties on this parameter), and a maximum mass $m_{\rm max}=50M_\odot$ which is motivated by models of initial BH mass spectra with pair-instability physics
in metal-poor environments 
(see e.g. Ref.~\cite{Spera:2018wnw}). 
Stellar evolution considerations and GW observations have been used to suggest that BHs may form  
with spin magnitudes as low as $\sim10^{-2}$~\cite{Fuller:2019sxi}. We assume all BHs to form with zero natal spin in our fiducial model ($\chi_0=0$), but this assumption does not affect the IMBH mass-spin relation, because the memory of the initial spin is quickly erased within a few merger generations. We compute the final mass and final spin 
using the expressions of Ref.~\cite{Gerosa:2016sys}, which are based on an interpolation between numerical relativity fits and the test-particle limit. 
Furthermore, the spin direction of the growing BH is isotropically sampled, modeling the stochasticity in the evolution of the BH spin as its mass grows through successive mergers in a roughly spherically symmetric environment like a cluster (the picture would be quite different in, e.g., an axisymmetric environment like a disk \cite{Yang:2019cbr,Santini:2023ukl}).

\begin{figure}
    \centering
    \includegraphics[width=0.49\textwidth]{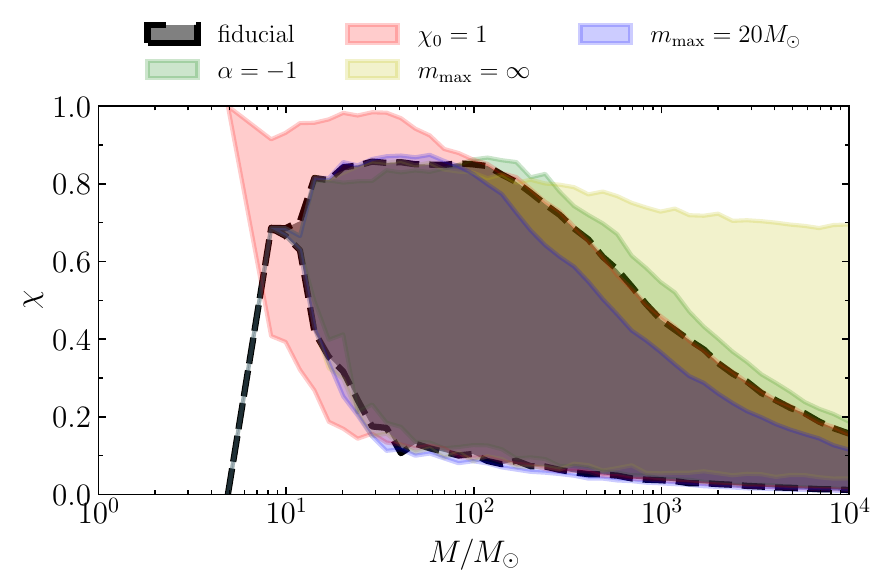}
    \caption{Areas encompassing $99\%$ of the realizations 
    for the mass-spin evolution of BHs through successive mergers assuming a seed mass of $5M_\odot$. The fiducial model for the BH mass spectrum (black filled region) assumes $(m_{\rm min}, m_{\rm max}, \alpha, \chi_0)=(5M_\odot, 50M_\odot, -2, 0)$. Variations from this model consider $\chi_0=1$ (red), $\alpha=-1$ (green), $m_{\rm max}=20M_\odot$ (blue), and $m_{\rm max}=\infty$ (yellow), while keeping all other parameters fixed to their fiducial value.}
    \label{fig:forward_modeling1}
\end{figure}

Figure~\ref{fig:forward_modeling1} shows the mass-spin evolution of IMBHs as they grow through repeated mergers. 
We denote by $M$ and $\chi$ the mass and spin of the growing BH, respectively. 
We illustrate the fiducial model described above, and then we vary each of the parameters ($\alpha, m_{\rm max}, \chi_0$) one at a time while keeping the other parameters fixed. 
As expected, the seed BH spin does not affect the mass-spin relation in the IMBH regime ($M>100M_\odot$), as seen by comparing 
the extreme cases where all seed BHs are nonspinning (black) and maximally spinning (red).
The model with $\alpha=-1$ (green) represents a shallower mass spectrum, which produces more heavy BHs (with mass close to $m_{\rm max}$) than the fiducial model.
In this scenario, at a fixed mass $M$, the spin has a higher value and takes longer to drop to zero.
By a similar argument, when we choose $m_{\rm max}=20M_\odot$ (blue), the BH spin magnitude decreases faster.
In the scenario where $m_{\rm max}=\infty$, the spin evolves the slowest because there is always a chance for the IMBH to pair up with a BH of similar mass, in which case the remnant spin will be around $\chi\sim0.7$.

\begin{figure}
    \centering
    \includegraphics[width=0.49\textwidth]{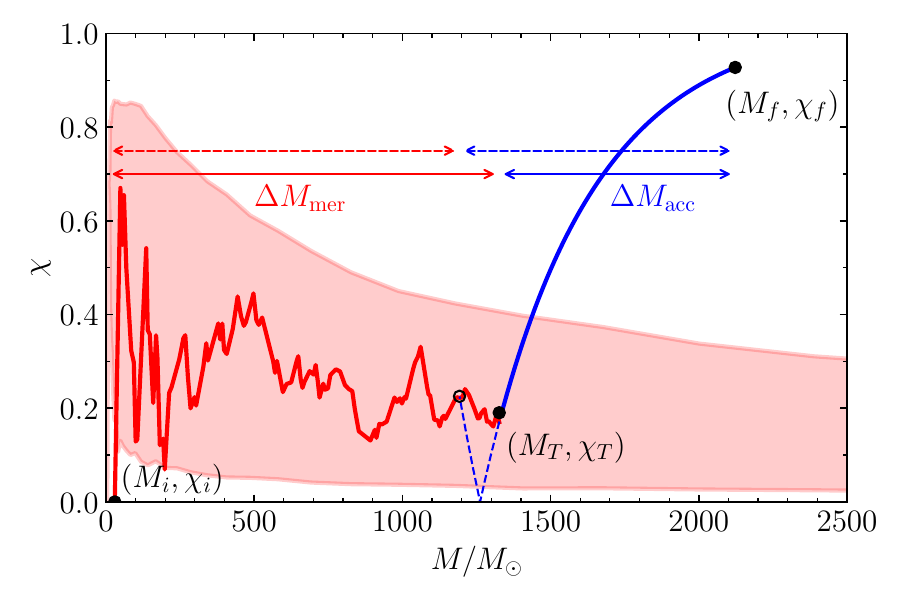}
    \caption{
    Mass-spin evolutionary history example for an IMBH growing by repeated BH mergers (red line, mass gained $\Delta M_{\rm mer}$) followed by a phase of coherent gas accretion (blue line, mass accreted $\Delta M_{\rm acc}$). The blue dashed branch corresponds to the case of a retrograde accretion phase that flips the spin direction. The solid red-filled region corresponds to the area that encompasses $99\%$ of realizations 
    of the fiducial hierarchical-merger model. 
    }
    \label{fig:forward_modeling2}
\end{figure}

We identify an exclusion region in the high-mass ($M\gtrsim500M_\odot$), high-spin ($\chi\gtrsim0.6$) region of the $M$--$\chi$ plane where the cumulative probability of an IMBH to be found with those parameters is $<0.5\%$.
We conclude that hierarchical mergers cannot efficiently cover this region,
and alternative mechanisms would be required to spin up the BH.
For extreme scenarios where $m_{\rm max}\to \infty$ and $\chi_0\simeq 1$, the region can be contaminated by hierarchical mergers. However these scenarios are not likely to be physical, because the extreme assumptions that either the BH mass spectrum extends to all masses or that all BHs have maximal spin are not supported by observations.

\subsection{Coherent gas accretion}
\label{sec:Coherent_gas_accretion}

Past their hierarchical growth, IMBHs can still increase their masses and spins through coherent gas accretion phases.
We assume that accretion occurs in the symmetry plane of the BH. While this is a simplification, the BH spin and the angular momentum of the disk tend to align because of the Lense-Thirring effect~\cite{1975ApJ...195L..65B}. Even if the disk and BH spin do not align efficiently, the required gas mass inferred under the assumption of alignment is a lower limit because, in the case of inclined accretion, a larger mass must be accreted to induce the same angular momentum variation projected onto the BH spin axis.

Let us assume that the forming IMBH grows up to some transition values $(M_{T}, \chi_{T})$  via hierarchical mergers, and then to some final values $(M_{f}, \chi_{f})$  via coherent accretion.  
For the latter part of the evolution we use standard prescriptions to compute the change in the IMBH's energy and angular momentum when it accretes a mass element orbiting at the ISCO~\cite{1970Natur.226...64B}. The relevant differential equation for the joint  evolution of $M$ and $\chi$ is
\begin{align}
    {d\chi \over dM} = {2\over3\sqrt{3}}{s\over M}{1 + 2\sqrt{3R(\chi, s) - 2}\over \left[ 1 - {2\over 3R(\chi, s)} \right]^{1/2}} - {2\chi\over M},
    \label{eq:dxdM}
\end{align}
where $R(\chi, s)$ is the ISCO radius normalized to the gravitational radius, and $s=1$ ($s=-1$) for prograde (retrograde) accretion. Equation~(\ref{eq:dxdM}) can easily be solved by numerical integration.
Notice that it does not involve the elapsed time during the IMBH growth; thus, we do not need to make assumptions about the gas density or the accretion rate.

The direction of $\chi_{T}$ and $\chi_{f}$ may differ if the IMBH undergoes an initial phase of retrograde accretion that flips the spin followed by a subsequent prograde accretion phase, in which the spin grows again.
In principle, it is impossible to know whether the spin of the IMBH flipped in the past just by measuring the spin magnitude $\chi_{f}$.
Thus, the two scenarios are degenerate, and in our fiducial forward modeling of BH growth, we give equal probability to both cases once the IMBH reaches the transition point. We also consider a case where we give more weight to the prograde scenario with probability $p_{\rm prog}=0.8$
(while illustrative, this specific choice is arbitrary; we are not aware of a reference or simulation that attempts to estimate the value of $p_{\rm prog}$).
In practice, this parameter can be inferred from measuring the mass, spin, and relative direction of spin and disk angular momentum for a population of accreting IMBHs. Naively, $p_{\rm prog}=N_{\rm prog}/N$, where $N_{\rm prog}$ is the number of IMBHs observed in their prograde accretion phase, and $N$ is the size of the population.
However, depending on the accretion rate, IMBHs may be more likely to be observed during their prograde phase, hence selection effects need to be taken into account when measuring $p_{\rm prog}$ from a population.

Figure~\ref{fig:forward_modeling2} shows an example of IMBH growth through repeated mergers (red curve) followed by a single gas accretion episode (blue curve), indicating both the prograde (solid) and retrograde (dashed) cases.
For retrograde accretion, the transition from repeated mergers to gas accretion needs to begin earlier (at the empty circle) and a larger amount of gas must be accreted compared to the prograde case (at the filled circle).
The IMBH's final mass $M_{f}$ thus results from the seed value $M_{i}$ with contributions from the repeated merger ($\Delta M_{\rm mer}$) and gas accretion ($\Delta M_{\rm acc}$) phases, respectively.
These parameters must satisfy the constraint 
\begin{equation}
M_{f}=M_{i} + \Delta M_{\rm mer} + \Delta M_{\rm acc}\,.
\label{eq:constraint}
\end{equation}

\begin{figure}
    \centering
    \includegraphics[width=0.49\textwidth]{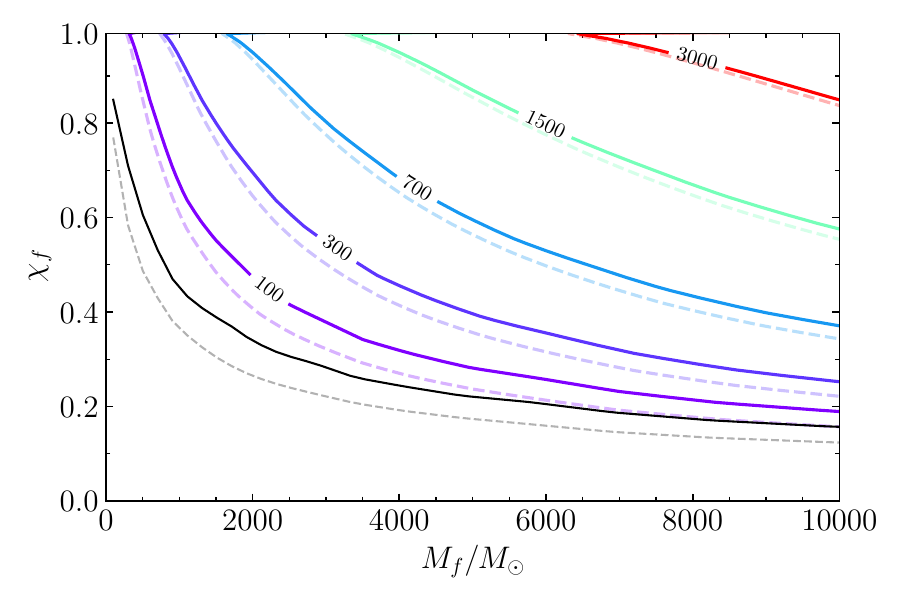}
    \caption{Lower bound $\Delta M_{\rm acc}^{\rm min}$ on the gas mass that must be accreted by an IMBH with observed mass $M_{f}$ and spin $\chi_{f}$. The black solid (dashed) line corresponds to the upper boundary of the area that encompasses the $99\%$ ($90\%$) of forward hierarchical-merger histories.}
    \label{fig:min_gas_accreted_99}
\end{figure}

\subsection{Minimum mass accreted}
\label{sec:Minimum_required_accreted}

Implementing the methodology above, we calculate the lower bound of the gas mass $\Delta M_{\rm acc}^{\rm min}$  that must be accreted by a putative IMBH observed in the forbidden region of the fiducial model.
Essentially, we integrate Eq.~\eqref{eq:dxdM} backward with a negative mass step ($dM<0$) from $(M_{f},\chi_{f})$ until $(M,\chi)$ reaches the upper boundary of the area encompassing $99\%$ ($90\%$) of hierarchical-merger realizations (cf.~Fig.~\ref{fig:forward_modeling2}).
If $M_{T}$ is the mass value of that boundary point, then $\Delta M_{\rm acc}^{\rm min}=|M_{f} - M_{T}|$, and we compute this on a fine grid of final conditions in the $M$--$\chi$ plane.
We show the results in Fig.~\ref{fig:min_gas_accreted_99}.
Most notably, we find that if $M_{f}\gtrsim1000M_\odot$ and $\chi_{f}\gtrsim0.6$, the IMBH must coherently accrete at least $\sim 100M_\odot$ of gas, while an IMBH with $(M_{f}, \chi_{f})\simeq(5000M_\odot, 0.9)$ must have accreted at least $\sim1500M_\odot$ of gas.
Combined with precise measurements of IMBH properties, these estimates can be used to constrain the amount of gas in the environment of highly spinning IMBHs.
In the next section, we infer the accreted gas mass, including, in particular, its most likely value and the upper bound on $\Delta M_{\rm acc}$.

\section{Distribution of accreted gas mass}
\label{sec:Distribution_of_accreted_gas_mass}

In this section, we infer the distribution of mass $\Delta M_{\rm acc}$ of accreted gas required to spin up the IMBH to its observed spin value, as well as the mass assembled by repeated mergers, $\Delta M_{\rm mer}$.
We assume that the mass $M_{f}$ and spin $\chi_{f}$ are measured with uncertainties $\delta M_{f}$ and $\delta \chi_{f}$, 
respectively, and that the IMBH lies in the ``forbidden'' (high-mass, high-spin) region which cannot be easily populated by hierarchical mergers.
We do not consider IMBHs whose parameters can be produced by repeated mergers because, in those cases, gas accretion is not required, and the simpler hierarchical scenario cannot be excluded.

To infer $\Delta M_{\rm mer}$ and $\Delta M_{\rm acc}$, we use the Markov chain Monte Carlo (MCMC) technique to explore different forward evolutionary histories that are consistent with the final observed IMBH parameters $(M_{f}, \chi_{f})$ accounting for some measurement errors $(\delta M_{f},\delta \chi_{f})$. 
We describe the computation of our likelihood function in Sec.~\ref{sec:Likelihood_function} below, and in Sec.~\ref{sec:MCMC_results} we present our inference results.

\subsection{Likelihood function}
\label{sec:Likelihood_function}

Our forward model is a stochastic map of the form 
\begin{align}
    F:X=(\Delta M_{\rm mer}, \Delta M_{\rm acc})\to Y=(M_{f}, \chi_{f}).
\end{align}
The stochasticity in this map arises from three latent random processes:
\begin{enumerate}[label=(\roman*)]
    \item The seed value $M_i$ is drawn from the initial BH mass spectrum.
    \item Spin directions are sampled isotropically, which induces randomness in the spin evolution during the hierarchical growth phase.
    \item The orbiting gas can be either prograde or retrograde; we assume probabilities of either $p_{\rm prog}=0.5$ or $p_{\rm prog}=0.8$  in favor of the former.
\end{enumerate}
We assume a Gaussian likelihood function  
\begin{align}
    \ln L(Y | X) = &-{1\over2}[Y - F(X)]^{\rm T}\Sigma^{-1}[Y - F(X)] \nonumber\\ &- {1\over2}\ln(4\pi^2 \det \Sigma),
\end{align}
where $\det$ denotes the determinant. We assume a diagonal covariance matrix ${\Sigma}={\rm diag}[\delta M_{f}^2, \delta\chi_{f}^2]$, 
ignoring correlations. This is a simplification: measurements of IMBH masses and spins may be correlated or anticorrelated as predicted, for example, by the so-called orbital hang-up effect in BH binary dynamics~\cite{Campanelli:2006uy}.
Our fiducial choice for the relative errors is $\delta M_{f}/M_{f}=\delta \chi_{f} / \chi_{f}=1\%$, but we also consider a more conservative value of $10\%$.
We choose a flat prior $\pi(X)$ in the range $[0,M_{f} + 10 \times \delta M_{f}]$ for both $\Delta M_{\rm mer}$ and $\Delta M_{\rm acc}$.
Finally, we sample the natural logarithm of the posterior $\ln p(X | Y) \propto \ln L(Y | X) + \ln\pi(X)$ and we explore the parameter space with the \textsc{emcee} package~\cite{2013PASP..125..306F}.
In Appendix~\ref{app:MCMC_diagnostics} we report some MCMC diagnostics of our runs.

\subsection{MCMC results}
\label{sec:MCMC_results}

\begin{figure*}
    \centering
    \includegraphics[width=0.495\textwidth]{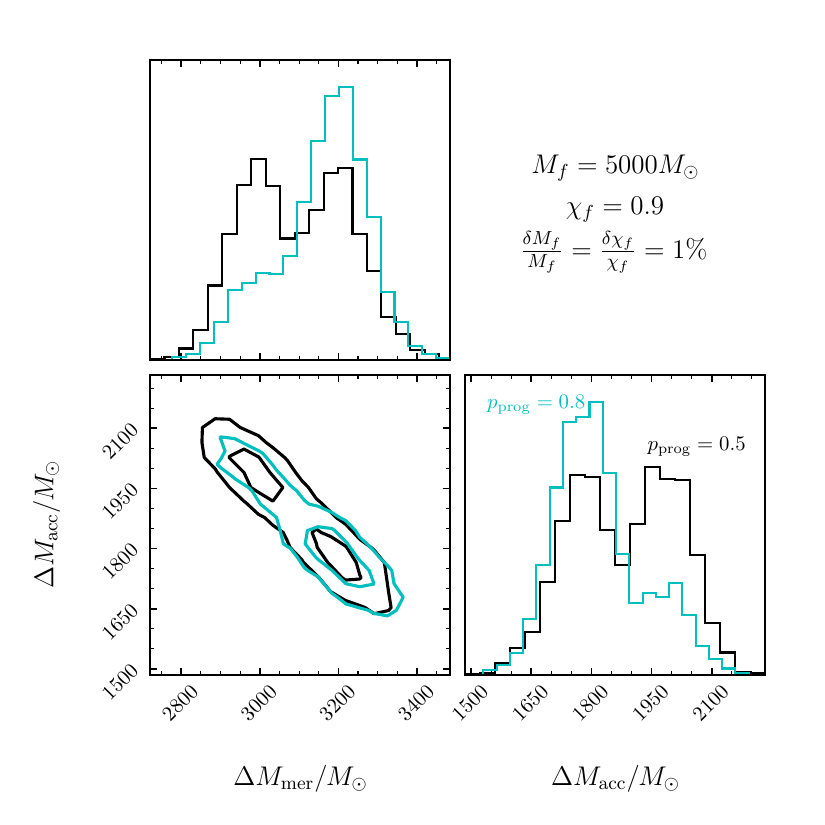}
    \includegraphics[width=0.495\textwidth]{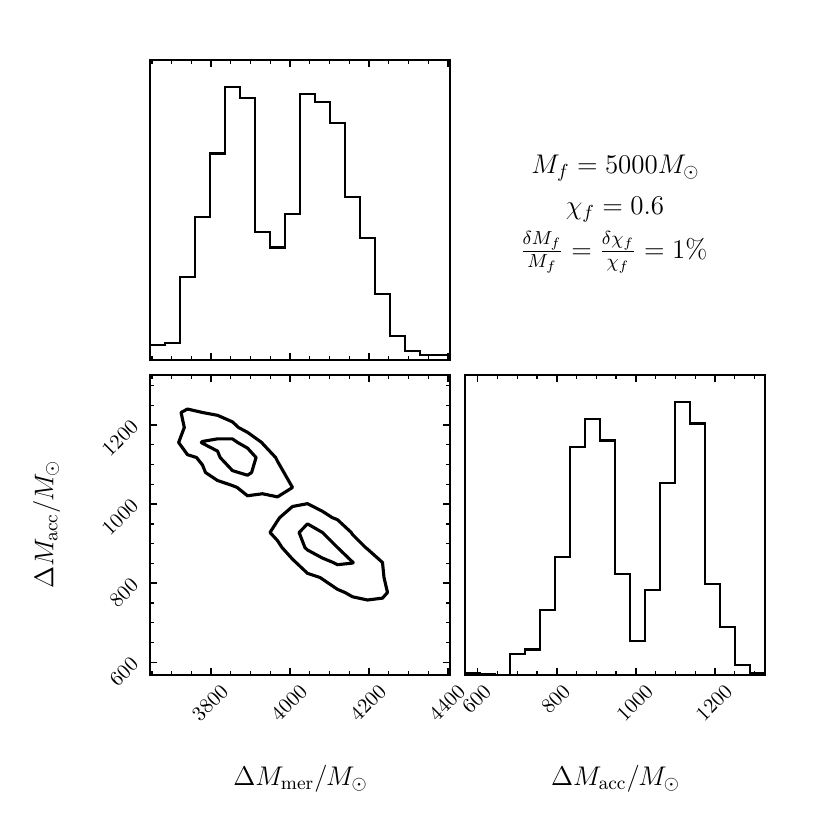}
    \includegraphics[width=0.495\textwidth]{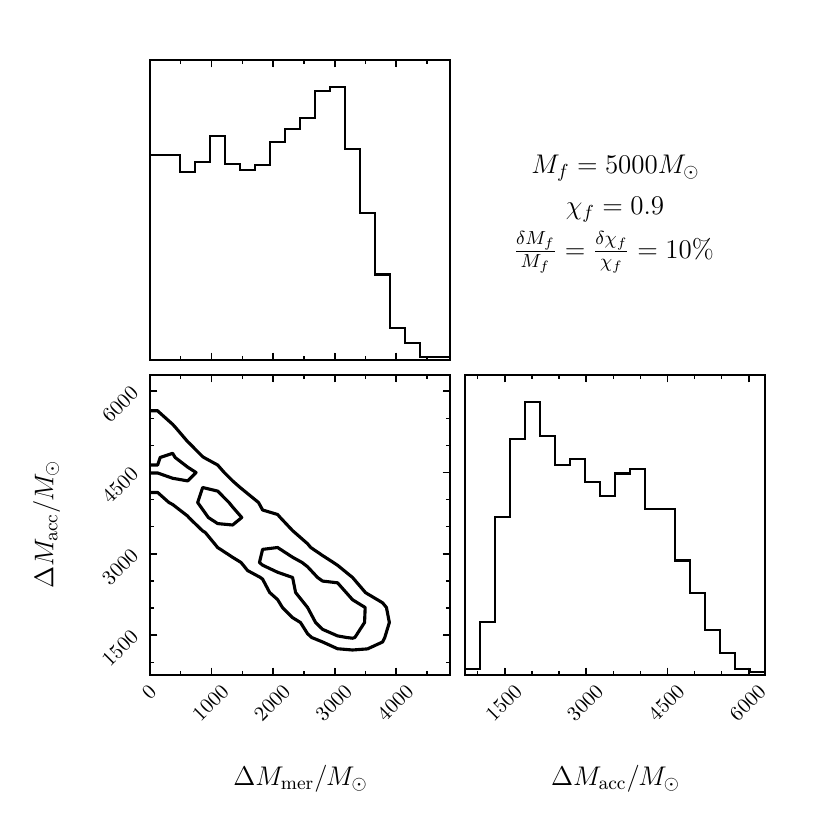}
    \includegraphics[width=0.495\textwidth]{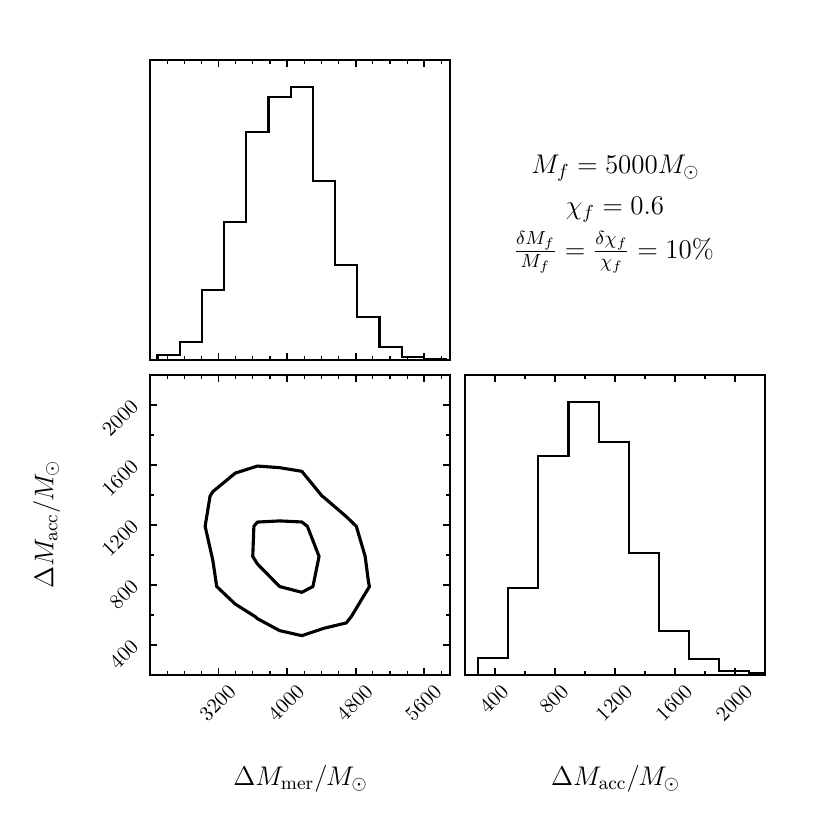}
    \caption{Posterior distribution inferred for $M_{f}=5000M_\odot$ and $\chi_{f}=0.9$ (left), $0.6$ (right), assuming fiducial hyperparameters. The relative measurement error on both $M_{f}$ and $\chi_{f}$ is fixed to 1\% (top) and 10\% (bottom). The 1-$\sigma$ (39\%) and 2-$\sigma$ (86\%) contours are shown. The probability for a purely prograde accretion phase $p_{\rm prog}$ is set to $0.5$; in the top left panel, we compare this against the case with $p_{\rm prog}=0.8$ (gray curves).}
    \label{fig:corner_5000}
\end{figure*}

\begin{figure}
    \includegraphics[width=0.49\textwidth]{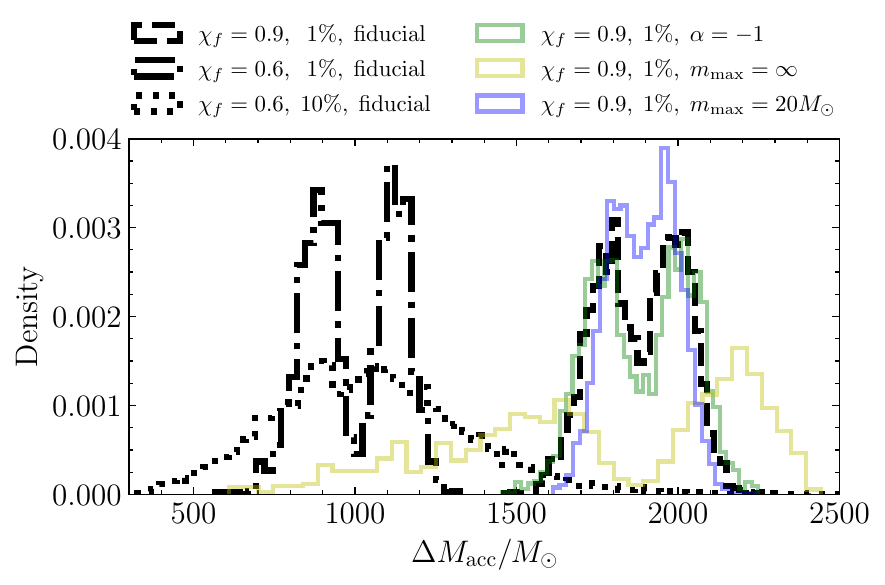}
    \caption{Marginalized posteriors on the accreted gas mass $\Delta M_{\rm acc}$ assuming different model parameters (cf. Fig.~\ref{fig:forward_modeling1}). The parameter $M_{f}$ is set to $5000M_\odot$.}
    \label{fig:DMacc_marginalized}
\end{figure}

In Fig.~\ref{fig:corner_5000} we show the posterior distribution for a putative IMBH detection with  $M_{f}=5000M_\odot$ and either $\chi_{f}=0.9$ (left column panels) or $\chi_{f}=0.6$ (right column panels) assuming the fiducial hyperparameters.
The top (bottom) row assumes a relative measurement error of $1\%$ ($10\%$) on the IMBH parameters.
Furthermore, in the top left panel, we compare $p_{\rm prog}=0.5$ (red) with $p_{\rm prog}=0.8$ (cyan).
In Appendix~\ref{app:Mf500} we discuss the case where $M_{f}=500M_\odot$.

The anticorrelation between $\Delta M_{\rm mer}$ and $\Delta M_{\rm acc}$ arises because of the constraint condition of Eq.~\eqref{eq:constraint}: if $M_{f}$ is well measured (as in the examples with error of $1\%$) and $M_{i}\ll M_{f}$, then the sum $\Delta M_{\rm mer} + \Delta M_{\rm acc}$ is roughly constant.
In addition, the distributions are clearly bimodal when the error is $1\%$. This is related to the degeneracy between the purely prograde case and the case where an initially retrograde spin changes direction under accretion. The purely prograde scenario corresponds to the lower-mass (higher-mass) peak in the $\Delta M_{\rm acc}$ ($\Delta M_{\rm mer}$) marginalized distribution, because a larger amount of gas must be accreted to flip the direction of the spin.
In comparison, when we give more weight to the purely prograde scenario (the $p_{\rm prog}=0.8$ case), the low-mass peak becomes more prominent.
The bimodality is less evident when the error is $10\%$ because the width of the peaks extends beyond their separation, so they blend together.

Near the peak of all posteriors, we have $\Delta M_{\rm acc}<\Delta M_{\rm mer}$. Generally, if a BH coherently accretes an amount of gas comparable to its mass, it spins up close to the extreme Kerr limit in a finite time~\cite{1970Natur.226...64B}.
Thus, a putative IMBH candidate should have coherently accreted an amount of gas $\lesssim M_{f}/2$; otherwise, its spin should be maximal, while we have assumed that the marginalized posterior on $\chi_{f}$ has vanishing support near unity when $\delta \chi_{f} / \chi_{f}=1\%$.
Therefore, in most of these examples the majority of the BH's mass is assembled through hierarchical mergers whenever $\chi_{f}$ is not maximal.
An exception is shown in the lower left panel of Fig.~\ref{fig:corner_5000}, where the posterior tail has to do with the fact that the distribution of $(M_{f},\chi_{f})$ rails against the maximal spin boundary.
A seed resulting in an IMBH with $\chi_{f}=1$ 
may have gained all of its mass through coherent gas accretion with no need for a contribution from BH mergers. In this case, the spin quickly approaches 1 from below, and the BH remains maximal until all of the available gas is accreted. Therefore, in this scenario, the posterior shows some support at the $\Delta M_{\rm mer}=0$ boundary, although most of the density is at $(\Delta M_{\rm mer}, \Delta M_{\rm acc})\approx(3000, 2000)M_\odot$.

Figure~\ref{fig:DMacc_marginalized} shows marginalized distributions of $\Delta M_{\rm acc}$ assuming $M_{f}=5000M_\odot$ and the different model variations illustrated in Fig.~\ref{fig:forward_modeling1}.
An IMBH with $\chi_{f}=0.9$ accretes more gas than an IMBH with $\chi_{f}=0.6$ by a factor of $\approx2$.
Additionally, the distance between the two peaks in each distribution is controlled by the width of the exclusion region around $\chi\sim0$~\cite{Gerosa:2021hsc}:
this width determines the length of the dashed blue line in Fig.~\ref{fig:forward_modeling2}, where the IMBH might undergo a phase of retrograde accretion.
In contrast, the width of the two modes
and of the overall distribution is caused by the variance in $\chi$ from stochasticity in the forward hierarchical-merger model at fixed mass $M$. 
Therefore, as $m_{\rm max}$ increases, the distributions become broader, and since the low-density region of forward hierarchical-merger histories 
widens, the peaks get further apart.
The same happens when we set $\alpha=-1$, mimicking the choice of a larger $m_{\rm max}$, although the effect is small.
Finally, as expected, the distribution broadens when the relative measurement error is raised to $10\%$, and the two peaks blend.

\section{Conclusions}
\label{sec:Conclusions}

Hierarchical mergers occurring in dense stellar clusters cannot efficiently produce IMBHs with both large masses ($M\gtrsim500M_\odot$) and large spins ($\chi\gtrsim0.6$). 
However, IMBHs may be later spun up in this forbidden region via coherent gas accretion.
This implies a map $(\Delta M_{\rm mer}, \Delta M_{\rm acc})\to (M_{f}, \chi_{f})$.
We inverted this map to constrain the required amount of gas, which depends on the measured IMBH parameters.  
For example, we found that if an IMBH is observed with $M_{f}=5000M_\odot$ and $\chi_{f}=0.9$ ($\chi_{f}=0.6$), then a mass of gas $\Delta M_{\rm acc}\approx 1800 M_\odot$ or $\approx 2000 M_\odot$ ($\approx 900 M_\odot$ or $\approx 1100 M_\odot$) must be accreted in a purely prograde or with an initially retrograde phase, respectively.
Thus, in order of magnitude, highly spinning IMBHs require roughly $\Delta M_{\rm acc}\lesssim M_{f}/2$ to be accreted.

This gas presumably exists in the cluster environment. It could originate from low-velocity stellar wind mass loss of giants in globular clusters and be continually replenished.
Deep radio observations have tentatively revealed the presence of neutral gas in some Milky~Way globular clusters~\cite{vanLoon:2005id}, and ionized hydrogen has also been detected in 47 Tucanae~\cite{Freire:2001wq,Abbate:2018pdf}, albeit in total amounts smaller than theoretical predictions~\cite{1975MNRAS.171..467T,vanLoon:2005id}.
Thus, the central gas reservoir cannot supply the thousands of solar masses required to significantly spin up any putative thousand-solar-mass IMBH in the core of globular clusters if IMBHs form through hierarchical mergers.
While dynamical signatures hint at the presence of IMBHs in the cores of some Milky~Way globular clusters \cite{2020ARA&A..58..257G,Huang:2024gpv,2024Natur.631..285H}, the absence of radio counterparts limits the mass to be $<1000M_\odot$ in most of these systems~\cite{Tremou:2018rvq}.

On the modeling side, the escape velocity of globular clusters is too low to allow for the hierarchical-merger products to be retained in the cluster (see, e.g., Ref.~\cite{Antonini:2018auk}, where the escape velocity needs to exceed $\approx300\,\rm km\, s^{-1}$ for an IMBH to be produced by hierarchical mergers).
Therefore, 
nuclear star clusters with escape velocity in the hundreds of $\rm km\, s^{-1}$ are more favorable environments for producing IMBHs through repeated mergers \cite{2021NatAs...5..749G}.
Perturbations on galactic scales may induce gas inflows into the central regions hosting the nuclear star cluster and the IMBH. This may lead to an episode of coherent gas accretion that could spin up the IMBH, as modeled here.
If these IMBHs are common in the nuclei of dwarf galaxies, their associated accretion signatures could be identified as low-luminosity active galactic nuclei (AGN)~\cite{2022NatAs...6...26R}.

If the gas-expulsion timescale is at least a few $\sim10\,\rm Myr$ 
in extremely dense 
nuclear star clusters (densities $>10^7M_\odot\,\rm pc^{-3}$),  hierarchical mergers and gas accretion processes may act simultaneously for as long as the gas reservoir is present deep into the BH subsystem~\cite{Kritos:2022non,Kritos:2024upo}.
Supermassive BHs may actually form in this way, and gas-rich nuclear star clusters are ideal environments for these processes, offering a possible formation route for the objects observed in the center of GN-z11~\cite{Maiolino:2023zdu}.

Our method is purposefully agnostic of any particular astrophysical scenario, although we envision the centers of dwarf galaxies with mini-AGN, such as NGC~4395~\cite{2003ApJ...588L..13F}, as potential environments where this scenario could be realized. A decoupling of the hierarchical BH merging and gas accretion processes could occur if gas inflow episodes drive large amounts of gas in the nuclear star cluster region of the dwarf galaxy at some later stage in the nuclear star cluster's evolution. In this paper, we have not attempted to carry out detailed simulations of these complex astrophysical processes. We note that, if the BH accretes at the Eddington rate, the accretion timescale is set by the Salpeter time $\tau_{S}\approx50\,\rm Myr$. If the time between successive mergers, written as the inverse of the merger rate $\Gamma_{m}^{-1}$, is larger than $\tau_{S}$ at the time of the gas inflow episode, then the two processes can be decoupled. To compute the merger rate, we write $\Gamma_{m}\approx n_{\rm BH}\sigma_{m} v$ where $n_{\rm BH}$, $\sigma_{m}$, and $v$ are the number density of stellar-mass BHs, merger cross section, and BH three-dimensional velocity dispersion, respectively. If we use the two-body capture cross section in Eq.~(4) of Ref.~\cite{Mouri:2002mc}, with $m_1\gg m_2$ the mass of the growing IMBH and a fixed $m_2=10M_\odot$, we obtain
\begin{align}
    \Gamma_{m}^{-1}&\approx 297{\,\rm Myr}\nonumber\\&\times \left(n_{\rm BH} \over {10^{5}\,{\rm pc}^{-3}}\right)^{-1}\left({m_1\over10^3M_\odot}\right)^{-12/7} \left({v\over 10\,\rm km\,s^{-1}}\right)^{-11/7}.
\end{align}
Thus, the condition $\Gamma_{m}^{-1}>\tau_{S}$ is satisfied for a measured IMBH mass of $\sim1000M_\odot$ in a typical environment with $v\sim10\,\rm km\,s^{-1}$, as long as the central BH number density $n_{\rm BH}$ does not significantly exceed $\sim10^6\,\rm pc^{-3}$.

Channels other than the scenario examined in this work may also contaminate the high-mass, high-spin region of the IMBH population.
A merger of two IMBHs with comparable masses will bring the spin to a high value ($\chi\sim0.7$) independent of the spin of the IMBH progenitors. Reference~\cite{Baibhav:2021qzw} shows that the remnant spin distribution for different progenitor spins and mass ratios
still displays an exclusion region above $\chi\sim0.8$, which we also identified in the $m_{\rm max}=\infty$ case in  Fig.~\ref{fig:forward_modeling1}.

Successive tidal disruption events may also contribute to forming IMBHs. In that case, the spin of the growing IMBH asymptotes to zero faster because the IMBH chaotically accretes smaller amounts of mass in each tidal disruption event compared to an IMBH growing via BH mergers. Effectively, this mimics much smaller values for $m_{\rm max}$ and $m_{\rm min}$.
Therefore, if tidal disruption events contributed to building an IMBH's mass, even larger amounts of gas should be accreted to reach high spin values.

Moreover, IMBHs can also form purely from gas accretion in gas-rich nuclear star clusters~\cite{Natarajan:2020avl}.
Depending on whether the accretion is coherent or chaotic,
the spin of the assembled IMBH evolves toward unity or asymptotes to zero, respectively. 
Finally, it is also possible that heavy BH seeds form with a high natal spin value~\cite{1975ARA&A..13..381E}. However, our current understanding of direct BH formation at those high masses is limited.

The symmetry of the environment is arguably our most crucial modeling assumption~\cite{Santini:2023ukl}. The IMBHs considered here grow hierarchically from mergers distributed isotropically, i.e., in spherical symmetry. Assuming spherical symmetry (axisymmetry) is an optimistic (conservative) choice to exclude large spin values, as repeated mergers tend to spin the IMBH down (up) after sufficiently long merger chains. Our results thus hold if the IMBH grows in a stellar cluster, where it is unreasonable to think that hierarchical encounters share a preferential direction. Assuming a cluster environment, our assumption of coherent disk accretion is then conservative, as it maximizes the spin gain. In other words, we can only constrain the \emph{minimum} accreted mass for IMBHs that grow in stellar clusters. For the same reason, our reasoning cannot be applied to IMBHs accreting in, e.g., AGN disks, where spin-up is expected by mergers alone. Furthermore, in those environments, the interplay between mass accretion and mergers is considerably more complex and cannot be easily split into separate phases, as done here.

Observations of GWs can yield measurements of the properties of IMBHs that do not depend on the complex physics of accretion and stellar dynamics.
While current ground-based GW detectors are not sensitive to masses $\gtrsim200M_\odot$~\cite{CalderonBustillo:2017skv,Chandra:2020ccy},
future observatories such as the Einstein Telescope~\cite{Punturo:2010zz}, Cosmic Explorer~\cite{Reitze:2019iox}, LISA~\cite{LISA:2017pwj}, and lunar experiments~\cite{Cozzumbo:2023gzs}
will be sensitive to the lower and upper end of the IMBH range, respectively~\cite{Reali:2024hqf,Ajith:2024mie,Chen:2022sae}.
Precise measurement of the masses and spins of the merging BHs could give us insights into their formation history and constrain the gas content of their nursery environments.

\begin{figure*}[t]
    \centering
    \includegraphics[width=0.47\textwidth]{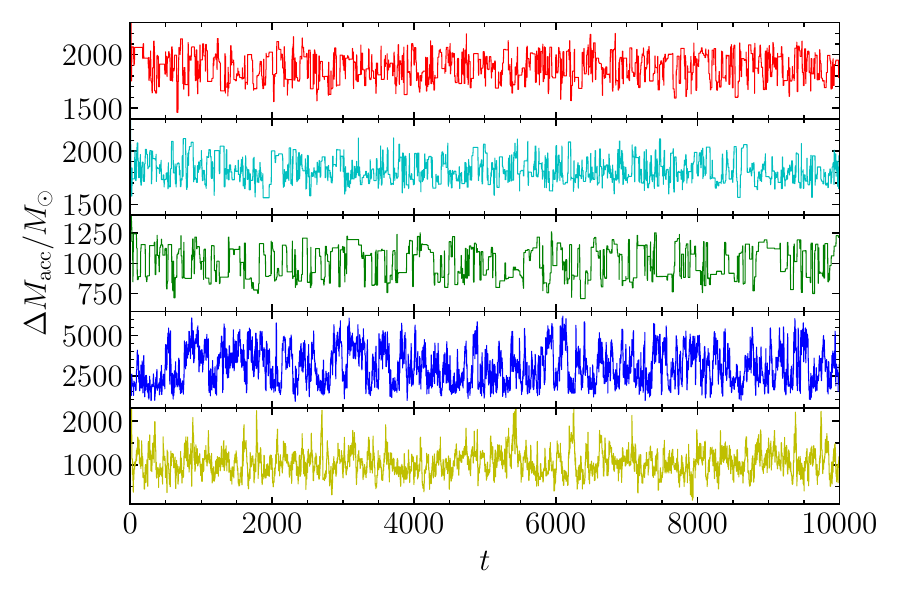}
    \includegraphics[width=0.47\textwidth]{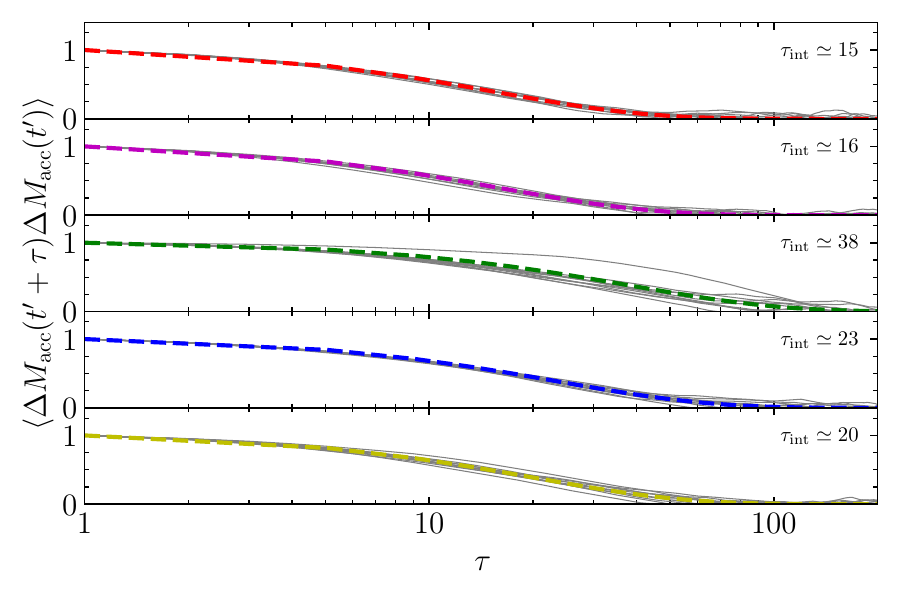}
    \caption{Left: trace plots of a single walker for each run presented in Fig.~\ref{fig:corner_5000}. Each color corresponds to a different case of Fig.~\ref{fig:corner_5000}: red (cyan) to the top left panel with $p_{\rm prog}=0.5$ ($p_{\rm prog}=0.8$), green to the top right, blue to the bottom left, and yellow to the bottom right. Right: the autocorrelation function of $\Delta M_{\rm acc}$ for all walkers. The dashed magenta lines represent the best-fit exponential model with the integrated autocorrelation time shown in the reported labels.}
    \label{fig:trace_and_autocorrelation}
\end{figure*}

\acknowledgments

K.K. is supported by the Onassis Foundation - Scholarship ID: F ZT 041-1/2023-2024. K.K., L.R. and E.B. are supported by NSF Grants No.~AST-2307146, No.~PHY-2207502, No.~PHY-090003, and No.~PHY-20043, by NASA Grants No.~20-LPS20-0011 and No.~21-ATP21-0010, by the John Templeton Foundation Grant 62840, by the Simons Foundation, and by the Italian Ministry of Foreign Affairs and International Cooperation Grant No.~PGR01167. 
This work was carried out at the Advanced Research Computing at Hopkins (ARCH) core facility~\cite{arch_link}, which is supported by the NSF Grant No. OAC-1920103.
D.G. is supported by 
ERC Starting Grant No.~945155--GWmining, 
Cariplo Foundation Grant No.~2021-0555, 
MUR PRIN Grant No.~2022-Z9X4XS, 
MUR Grant ``Progetto Dipartimenti di Eccellenza 2023-2027'' (BiCoQ),
and the ICSC National Research Centre funded by NextGenerationEU. 
Computational work was performed at CINECA with allocations 
through INFN and Bicocca.

\section*{Data availability}

No data were created or analyzed in this study.

\appendix

\section{MCMC diagnostics}
\label{app:MCMC_diagnostics}

The left panel of Fig.~\ref{fig:trace_and_autocorrelation} shows the sampled values as a function of step number (i.e., the trace plot) of $\Delta M_{\rm acc}$ for a single walker for the cases shown in Fig.~\ref{fig:corner_5000}. 
Our production runs make use of 10 MCMC walkers, which present qualitatively similar properties. 
Based on our physical expectation that the distributions we are sampling should be slightly bimodal, we use a combination of the {\tt DEMove} and {\tt DESnookerMove} with weights 0.8 and 0.2, respectively, as suggested in the \textsc{emcee} documentation. 
With this choice, the sampler explores each peak locally and occasionally jumps to the other peak.
A standard heuristic measure of convergence for an MCMC sampler is the integrated autocorrelation time, denoted by $\tau_{\rm int}$~\cite{Hogg:2017akh}.
We estimate this by fitting an exponential model of the form $\exp[-(t-1)/\tau_{\rm int}]$ to the normalized autocorrelation function for each walker and then take the mean inferred value from all fits.
We compute the autocorrelation function for $\Delta M_{\rm acc}$ as in Ref.~\cite{2013PASP..125..306F}. The results are shown in the right panel of Fig.~\ref{fig:trace_and_autocorrelation} along with the estimated integrated autocorrelation values.
We thin the chains of each walker by $\tau_{\rm int}$ so that our draws represent quasi-independent samples.

\begin{figure}
   \centering
   \includegraphics[width=0.49\textwidth]{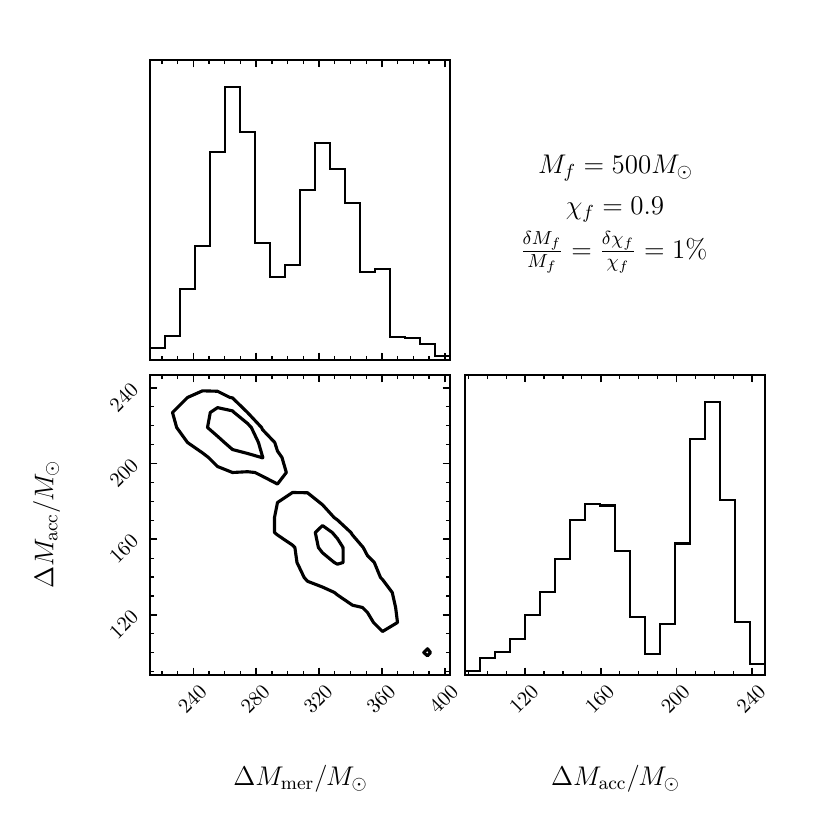}
   \caption{Same as the top left panel of Fig.~\ref{fig:corner_5000}, assuming $p_{\rm prog}=0.5$ but with $M_{f}=500M_\odot$.}
   \label{fig:corner_500}
\end{figure}

\section{$M_{f}=500M_\odot$, $\chi_{f}=0.9$}
\label{app:Mf500}

In this appendix, we quantify through an example the impact of changing the measured IMBH mass $M_{\rm f}$.

In Fig.~\ref{fig:corner_500} we repeat the inference assuming $(M_f,\chi_f)=(500M_\odot,0.9)$ along with the fiducial parameters.
In this case, the IMBH should have accreted $\approx150M_\odot$ or $\approx210M_\odot$, depending on whether the spin flipped direction during the accretion phase. The posterior is bimodal because we cannot distinguish between those two scenarios, as we have set $p_{\rm prog}=0.5$ in this example.

\bibliography{Gas_accreted_by_IMBHs}

\end{document}